\documentclass[]{aa} 

\usepackage{graphicx}
\usepackage{txfonts}

\begin{document}
\title{Investigating the possible connection between $\lambda$ Bootis stars and
intermediate Population\,II type stars}

\author{E.~Paunzen\inst{1}
\and I.Kh.~Iliev\inst{2}
\and L.~Fossati\inst{3}
\and U.~Heiter\inst{4}
\and W.W.~Weiss\inst{5}}
\institute{Department of Theoretical Physics and Astrophysics, Masaryk University,
Kotl\'a\v{r}sk\'a 2, 611\,37 Brno, Czech Republic \\
\email{epaunzen@physics.muni.cz}
\and
Rozhen National Astronomical Observatory, Institute of Astronomy of the Bulgarian 
Academy of Sciences, P.O. Box 136, BG-4700 Smolyan, Bulgaria
\and
Argelander-Institut f{\"u}r Astronomie der Universit{\"a}t Bonn, Auf dem H{\"u}gel 71, 53121, Bonn, Germany
\and
Institutionen f{\"o}r fysik och astronomi, Uppsala universitet, Box 516, 751 20, Uppsala, Sweden
\and
Institute for Astrophysics, University of Vienna, T{\"u}rkenschanzstr. 17, 1180 Vienna, Austria}

   \date{} 
  \abstract
{The $\lambda$ Bootis stars are located at the upper main sequence of the H--R
diagram and exhibit a peculiar abundance pattern. The light elements (C, N, O,
and S) present solar abundances whereas all other elements are moderately to
strongly underabundant. It has not yet been determined whether that abundance
pattern is intrinsic, or is restricted to the stellar surface.}  
{If we follow the hypothesis that the $\lambda$ Bootis stars are intrinsically
metal-weak, then there should be a connection with the intermediate
Population II and F-weak objects.  Such a possible affinity has not been
previously investigated.}  
{We present detailed elemental abundances, including those of the light
elements carbon and oxygen, for 38 bright intermediate Population II and F-weak
objects. In addition, we investigate the kinematic characteristics of the
groups.}  
{From photometric, spectroscopic, and kinematic data, there is no distinction
between the intermediate Population II and F-weak type stars. We therefore
conclude that the two groups are identical. However, it is possible to
distinguish the $\lambda$ Bootis stars from the intermediate Population II
stars on the basis of elemental abundances, though not in terms of their 
kinematics.}  
{The $\lambda$ Bootis stars seem to be distinct from the intermediate
Population II group. Further asteroseismologic investigations and analyses of
spectroscopic binary systems are needed to strengthen this conclusion.}

\titlerunning{$\lambda$ Bootis stars and intermediate Population\,II type 
stars}
\authorrunning{Paunzen et al.}
\maketitle

\section{Introduction}

The $\lambda$ Bootis stars are a small group (only 2\%) of late-B to early-F
stars that show moderate to extreme surface underabundances (up to a factor of
100) of most Fe-peak elements, but solar abundances of lighter elements (C, N,
O, and S).  They stand out among the chemically peculiar stars of
the upper main sequence because all other groups exhibit strong overabundances
of elements, probably caused by  magnetic fields, slow rotation or
atmospheric diffusion \citep[e.g. ][]{Neto08}.  Several members of the group
exhibit a strong infrared excess and a disk \citep{Paun03,Boot13}.

To explain the $\lambda$ Bootis phenomenon, \citet{Venn90} suggested that the
peculiar chemical abundance is caused by selective accretion of circumstellar
material. One of the principal features of that hypothesis is that the observed
abundance anomalies are restricted to the stellar surface.  Members of the
group are therefore true Population I type objects.  Later, \citet{Kamp02} and
\citet{Mart09} developed models that describe the interaction of the star with
its local interstellar and/or circumstellar environment, whereby different
degrees of underabundance are produced by different amounts of accreted
material relative to the photospheric mass.  The fact that the fraction of
$\lambda$ Bootis stars on the main sequence is so small would then be a
consequence of the low probability of a star-cloud interaction, and by the
effects of meridional circulation which dissolves any accretion pattern a few
million years after the accretion has stopped.  The hot cut-off would be the
limitation of significant stellar winds for stars with
$T_\mathrm{eff}$\,$>$\,12\,000\,K whereas the cool end, at about 6500\,K, would
be defined by convection which prevents the accreted material from appearing at
the stellar surface.

\begin{table*}[ht]
\caption{The astrophysical and kinematic parameters for our target sample.
The uncertainties in the final digits of the corresponding quantity are given in parentheses.}
\label{parameters}
\begin{center}
\begin{tabular}{llccccccccc}
\hline
\hline 
HD & HIP & $V$ & \ensuremath{T_{\mathrm{eff}}} & \ensuremath{\log g} & \ensuremath{M_{\mathrm{V}}} & \ensuremath{{\upsilon}\sin i} &  
\ensuremath{R_{\mathrm{V}}} & $U$ & $V$ & $W$ \\
& & [mag] & [K] & [dex] & [mag] & [km\,s$^{-1}$] & [km\,s$^{-1}$] & [km\,s$^{-1}$] & [km\,s$^{-1}$] & [km\,s$^{-1}$] \\ 
\hline
693	&	910	&	4.91	& 6200(55)	&	4.10(7)	&	3.53(2)	&	6.0(3)	&	+13.5(4.1)	&	+18.9(2)	&	$-$13.4(1.1)	&	$-$16.8(4.0)	\\
3229	&	2787	&	5.94	&	6350(88)	&	3.80(15)	&	2.22(5)	&	6.0(6)	&	+6.2(2.5)	&	$-$25.5(9)	&	$-$29.2(1.1)	&	$-$15.2(2.3)	\\
4813	&	3909	&	5.18	&	6150(34)	&	4.35(11)	&	4.23(2)	&	5.4(3)	&	+9.3(2.2)	&	+21.2(4)	&	$-$2.3(6)	&	$-$13.6(2.1)	\\
7476	&	5833	&	5.70	&	6450(50)	&	3.90(15)	&	2.48(6)	&	6.8(1)	&	+25.5(2.5)	&	$-$26.7(1.0)	&	+41.9(1.2)	&	$-$4.2(2.3)	\\
9919	&	7535	&	5.60	&	6850(49)	&	4.00(5)	&	2.94(7)	&	105.9(7.8)	&	$-$1.0(2.5)	&	+9.2(1.3)	&	+8.0(1.1)	&	$-$0.4(1.9)	\\
12414	&	9473	&	7.06	&	6250(47)	&	4.25(15)	&	3.72(5)	&	6.0(9)	&	+14.1(2.5)	&	$-$22.7(1.4)	&	$-$20.5(9)	&	$-$9.9(2.0)	\\
13555	&	10306	&	5.23	&	6450(85)	&	4.15(11)	&	2.80(5)	&	7.4(4)	&	+6.4(6)	&	$-$20.7(6)	&	$-$11.8(5)	&	+3.7(4)	\\
16626	&	12656	&	7.02	&	6550(99)	&	4.20(15)	&	3.32(8)	&	17.0(2.5)	&	+14.9(2.5)	&	$-$17.6(1.8)	&	+3.8(1.8)	&	$-$13.5(6)	\\
17948	&	13665	&	5.59	&	6450(65)	&	4.20(8)	&	3.48(3)	&	9.0(1.4)	&	+29.2(2.5)	&	$-$30.9(1.8)	&	+9.0(1.7)	&	+12.9(2)	\\
24740	&	18471	&	5.70	&	6900(81)	&	3.75(1)	&	2.42(8)	&	21.0(2.1)	&	+32.2(2.5)	&	$-$31.4(2.2)	&	$-$22.3(8)	&	$-$19.2(1.3)	\\
25457	&	18859	&	5.36	&	6250(59)	&	4.45(24)	&	3.94(4)	&	19.2(1.2)	&	+18.1(8)	&	$-$8.3(6)	&	$-$28.8(4)	&	$-$12.2(6)	\\
26015	&	19261	&	6.02	&	6750(43)	&	4.15(12)	&	2.62(9)	&	30.0(4.5)	&	+37.7(2.7)	&	$-$42.7(2.4)	&	$-$20.6(8)	&	$-$0.7(1.5)	\\
27397	&	20219	&	5.58	&	7300(57)	&	3.95(5)	&	2.30(8)	&	100.0(5.8)	&	+42.0(2.5)	&	$-$44.8(2.3)	&	$-$19.2(6)	&	$-$2.8(1.3)	\\
27524	&	20349	&	6.79	&	6550(87)	&	4.30(14)	&	3.22(10)	&	90.0(13.5)	&	+37.1(2.5)	&	$-$41.4(2.3)	&	$-$19.8(8)	&	$-$0.1(1.3)	\\
27808	&	20557	&	7.13	&	6200(61)	&	4.35(21)	&	4.05(8)	&	12.0(1.8)	&	+36.8(2.9)	&	$-$40.4(2.7)	&	$-$18.4(7)	&	$-$1.2(1.2)	\\
29391	&	21547	&	5.22	&	7300(73)	&	4.20(7)	&	2.84(4)	&	77.9(1.8)	&	+21.0(2.5)	&	$-$14.0(2.0)	&	$-$16.2(7)	&	$-$10.1(1.3)	\\
33256	&	23941	&	5.12	&	6350(80)	&	3.95(21)	&	3.11(4)	&	9.9(6)	&	+10.2(1.2)	&	$-$10.4(1.0)	&	$-$6.3(5)	&	+1.8(5)	\\
65123	&	38870	&	6.35	&	6300(41)	&	4.25(17)	&	2.44(13)	&	13.0(2.0)	&	+1.1(2.1)	&	$-$22.9(2.0)	&	+9.7(1.5)	&	$-$38.3(2.4)	\\
69897	&	40843	&	5.15	&	6300(47)	&	4.35(8)	&	3.85(3)	&	4.2(2)	&	+32.8(5)	&	$-$24.3(4)	&	$-$38.5(5)	&	+7.2(3)	\\
110093	&	61396	&	7.23	&	7450(36)	&	4.00(15)	&	2.48(9)	&	115.0(20)		&	$-$12.0(2.5)	&	$-$7.4(1.3)	&	$-$13.0(1.8)	&	$-$11.9(1.4)	\\
114642	&	64407	&	5.03	&	6350(59)	&	3.90(16)	&	2.46(3)	&	13.5(6)	&	$-$14.1(2.5)	&	+22.4(1.2)	&	$-$11.7(1.4)	&	$-$41.8(1.8)	\\
125111	&	69735	&	6.20	&	6700(72)	&	4.15(1)	&	2.78(6)	&	9.5(3)	&	$-$25.0(1.2)	&	$-$25.9(6)	&	$-$28.5(7)	&	$-$13.2(1.2)	\\
125276	&	69965	&	5.87	&	6100(88)	&	4.55(18)	&	4.62(3)	&	3.9(3)	&	$-$21.0(2.5)	&	$-$41.5(1.8)	&	+9.0(1.3)	&	+22.3(1.4)	\\
128167	&	71284	&	4.47	&	6750(61)	&	4.30(4)	&	3.52(2)	&	9.5(1)	&	+0.2(2)	&	+2.1(1)	&	+15.9(1)	&	$-$5.2(2)	\\
135208	&	74593	&	6.77	&	6450(79)	&	3.85(15)	&	2.58(9)	&	9.0(1.4)	&	$-$24.9(7)	&	$-$36.7(9)	&	$-$21.7(9)	&	$-$0.3(1.0)	\\
137006	&	75342	&	6.11	&	7450(62)	&	4.05(6)	&	2.41(9)	&	127.8(7.4)	&	$-$2.2(2.5)	&	+12.0(1.9)	&	+7.7(6)	&	$-$15.9(1.8)	\\
147449	&	80179	&	4.83	&	7000(49)	&	4.05(5)	&	2.63(6)	&	77.7(6)	&	$-$49.8(2.5)	&	$-$49.7(2.1)	&	$-$18.8(6)	&	$-$10.2(1.4)	\\
157950	&	85365	&	4.53	&	6700(43)	&	4.00(69	&	2.13(4)	&	27.0(3)	&	$-$0.1(9)	&	+1.4(8)	&	$-$12.0(4)	&	+7.9(4)	\\
185124	&	96556	&	5.45	&	6650(43)	&	4.15(19)	&	2.88(3)	&	84.3(6.1)	&	$-$37.6(2.5)	&	$-$33.6(2.0)	&	$-$22.8(1.4)	&	$-$9.4(6)	\\
189712	&	98565	&	7.46	&	6450(120)	&	3.85(15)	&	2.85(13)	&	8.0(1.2)	&	+31.5(2.5)	&	+32.0(1.7)	&	+13.9(1.9)	&	$-$0.6(9)	\\
191841	&	99558	&	7.33	&	6450(72)	&	4.25(10)	&	2.72(15)	&	14.0(2.1)	&	+6.6(2.5)	&	+9.8(2.0)	&	$-$7.9(1.6)	&	$-$9.0(1.1)	\\
198390	&	102805	&	5.90	&	6500(75)	&	4.30(11)	&	3.51(3)	&	6.7(2)	&	+2.3(2.5)	&	$-$11.9(1.2)	&	+10.4(2.0)	&	+0.8(8)	\\
199124	&	103261	&	6.56	&	7200(58)	&	3.80(9)	&	2.03(9)	&	159.1(11.5)	&	$-$6.6(4.0)	&	$-$11.7(2.4)	&	+3.2(2.6)	&	+3.4(1.9)	\\
208362	&	108147	&	7.46	&	6750(116)	&	4.15(15)	&	3.20(10)	&	17.9(1.4)	&	+20.6(4.0)	&	$-$21.3(7)	&	+19.5(4.0)	&	+3.7(8)	\\
211575	&	110091	&	6.40	&	6500(39)	&	4.15(13)	&	3.31(5)	&	19.0(1.6)	&	+15.8(1.4)	&	+17.0(5)	&	+3.5(1.0)	&	$-$11.4(1.0)	\\
218804	&	114430	&	6.00	&	6400(81)	&	4.10(9)	&	3.73(3)	&	17.2(9)	&	$-$43.4(2.5)	&	+46.0(7)	&	$-$35.6(2.3)	&	$-$2.6(7)	\\
222368	&	116771	&	4.13	&	6200(36)	&	4.15(15)	&	3.43(2)	&	6.3(1)	&	+4.1(3.5)	&	$-$7.8(3)	&	$-$27.1(2.1)	&	$-$25.3(2.8)	\\
223346	&	117445	&	6.47	&	6350(71)	&	3.90(13)	&	2.76(7)	&	18.9(1.4)	&	$-$25.0(2.5)	&	+3.2(4)	&	$-$21.3(1.4)	&	+15.9(2.1)	\\
\hline
\end{tabular}
\end{center}
\end{table*}

The important question as to whether or not $\lambda$ Bootis stars are
intrinsically metal-weak does not seem to have been raised very often in the
literature. Recently, \citet{Casa09} made a detailed asteroseismological
analysis of the pulsational behaviour of 29~Cygni (HR~7736), which is one of
the prototypes of the $\lambda$ Bootis group and exhibits very strong
underabundances of the Fe-peak elements compared to the Sun
($\left[\mathrm{Fe/H}\right]$\,=\,$-$1.63\,dex; Paunzen et al. 2002).  From
their best-fitting models they concluded that 29 Cyg is an intrinsic metal-weak
main-sequence object. Unfortunately, there is only one other similar analysis
in the literature: \citet{Breg06} investigated the pulsational characteristics
of HD~210111 (HR~8437) and found no differences from the behaviour of the
classical $\delta$ Scuti star, FG Virginis, which has a similar evolutionary
status. The star HD~210111 was later found to be a spectroscopic binary system with
almost equal components \citep{Paun12}. Although the components have
[M/H]\,=\,$-$1.0\,dex, the pulsational models of \citet{Breg06} adopted solar
abundances.  It would be very interesting to re-analyse their observations with
corresponding metal-weak models.

Incorporating unrepresentative 
metal abundances in the models 
has consequences not only for the theories already developed to explain 
the $\lambda$ Bootis phenomenon, but also for 
the calibration of the astrophysical parameters.
For example, in order to derive stellar ages one would need to use
isochrones calculated for abundances that were intrinsically non-solar.

As a rider to that hypothesis, it would be difficult to describe the $\lambda$
Boo stars as true Population I type objects.  A first possible connection
between these star groups was suggested by \citet{Gray89}, who introduced a new
classification scheme for the intermediate Population II (hereafter ip\,II) stars.  In this paper we present a fresh
detailed investigation of a possible connection between the F-weak objects and ip\,II ones, using detailed elemental
abundances, photometric diagrams, and kinematic data.

\section{Target selection, observations, and reductions}

We chose 38 bright ($V$\,$<$\,7.5\,mag) F-weak and ip\,II stars from the
samples by \citet{Jasc89} and \citet{Gray89}; all have parallaxes measured by
{\it Hipparcos} \citep{Vanl09}.  We expressly selected objects that overlap
the $\lambda$ Bootis stars in terms of effective temperature and surface gravity.
The stars are listed in Table \ref{parameters}.

Observations of the sample were made between 2001--2002 with the coud\'{e}
spectrograph of the 2m Ritchey-Chr\'etien telescope of the Bulgarian National
Astronomical Observatory, Rozhen, using the Photometrics AT200 camera with a
SITe SI003AB 1024\,$\times$\,1024 CCD chip. Typical
seeing was 2--3\,$\arcsec$; the slit width was set to 300\,$\mu$m, which
corresponded to 0$\farcs$8 on the sky.  A Bausch \& Lomb grating with 632
lines\,mm$^{-1}$ was centred on two different spectral regions,
7120\,\AA\,(C-region) and 7770\,\AA\,(O-triplet), and provided a resolving
power of about 30,000 and 22,000, respectively. A hollow-cathode ThAr lamp
produced a reference spectrum of comparison lines with a FWHM of about 2
pixels.  A tungsten projection lamp mounted in front of the entrance slit
provided flat-fielding.

Additional high S/N ratio spectra were also obtained during several observing
runs in 2001 and 2002 with the Sandiford Cassegrain Echelle Spectrograph
\citep{McCa93} on the 2.1-m telescope at McDonald Observatory, USA.  Those
spectra covered continuously a wavelength range from about 4800 to 7000\,\AA\
with a resolving power of about 60,000.  Each night the spectrum of a
broad-lined B star was obtained with a S/N ratio that exceeded those of the
programme stars by a factor of several, to enable corrections for telluric lines
where necessary.

By using IRAF\footnote{IRAF is distributed by the National Optical Astronomy
Observatories, which are operated by the Association of Universities for
Research in Astronomy, Inc., under cooperative agreement with the National
Science Foundation.} we processed the CCD images, subtracted scattered light,
extracted echelle orders, and performed wavelength calibrations and
continuum normalization.
Equivalent widths were measured from the normalized spectra by adopting a
Gaussian approximation to the line profile.

Photometric data were taken from the General Catalogue of Photometric Data
(GCPD\footnote{http://obswww.unige.ch/gcpd/}).  Where possible, averaged and
weighted mean values were used throughout.
To derive the effective temperatures and surface gravities, we employed the
following calibrations for the individual photometric systems:

\begin{itemize}
\item Johnson {\it UBV}: \citet{Napi93};
\item Str{\"o}mgren $uvby\beta$: \citet{Moon85,Napi93};
\item Geneva 7-colour: \citet{Kobi90,Kuen99}.
\end{itemize}

When parameters from several different systems and calibrations were derived, a
simple average and standard deviation was calculated; these values are
listed in Table \ref{parameters}. The average uncertainties were
$\sigma\overline{T_{\mathrm{eff}}}$\,=\,63(23)\,K and $\sigma\overline{\log
g}$\,=\,0.12(6)\,dex.  No significant outliers were detected, so all the
photometric measurements are intrinsically consistent.

\begin{table*}
\caption{Element abundances for our sample of stars, relative to the solar 
values given by \citet{Grev96}. Typical uncertainties are $\pm$0.2\,dex.}
\label{abundances}
\begin{center}
\begin{tabular}{lcccccccccc}
\hline
\hline 
HD	&	\ion{C}{I}	&	\ion{Si}{I}	&	\ion{Ti}{I}	&	\ion{Cr}{I}	&	\ion{Fe}{I} 	&	\ion{Ni}{I}	&	\ion{K}{I}	&	\ion{Ca}{I}	&	\ion{O}{I} (LTE)	&	\ion{O}{I} (NLTE)	\\
\hline
693	&	$-$0.10	&	+0.12	&	$-$0.25	&	$-$0.20	&	$-$0.36	&	$-$0.38	&		&		&		&		\\
3229	&	$-$0.02	&	+0.02	&	$-$0.19	&	$-$0.03	&	$-$0.23	&	$-$0.31	&		&		&		&		\\
4813	&	+0.17	&		&	$-$0.32	&	$-$0.06	&	$-$0.17	&	$-$0.23	&		&		&		&		\\
7476	&	$-$0.05	&	$-$0.08	&	$-$0.11	&	+0.05	&	$-$0.19	&	$-$0.28	&		&		&		&		\\
9919	&	$-$0.52	&	+0.06	&		&		&	$-$0.11	&	+0.78	&	+0.78	&	+0.33	&	+0.75	&	+0.09	\\
12414	&	$-$0.23	&	$-$0.14	&	$-$0.25	&	$-$0.22	&	$-$0.34	&	$-$0.40	&	$-$0.12	&	$-$0.24	&	+0.14	&	$-$0.38	\\
13555	&	$-$0.15	&	$-$0.12	&	$-$0.06	&	+0.08	&	$-$0.20	&	$-$0.27	&	+0.58	&	+0.05	&	+0.19	&	$-$0.37	\\
16626	&	$-$0.29	&	$-$0.44	&	$-$0.14	&	$-$0.18	&	$-$0.37	&	$-$0.34	&	+0.16	&	+0.19	&	+0.17	&	$-$0.42	\\
17948	&	$-$0.29	&	$-$0.21	&		&		&	$-$0.31	&	$-$0.50	&	$-$0.05	&	$-$0.15	&	+0.17	&	$-$0.40	\\
24740	&	$-$0.14	&	$-$0.04	&		&		&	$-$0.09	&	$-$0.27	&	+1.10	&	+0.69	&	+1.12	&	+0.44	\\
25457	&	$-$0.02	&	+0.15	&		&		&	+0.20	&	$-$0.14	&	+0.47	&	+0.52	&	+0.33	&	$-$0.18	\\
26015	&	+0.23	&	+0.12	&		&		&	+0.13	&	+0.04	&	+0.88	&	+0.80	&	+0.86	&	+0.23	\\
27397	&	+0.03	&	$-$0.36	&		&		&	$-$0.06	&	+1.49	&	+0.44	&	+1.14	&	+1.06	&	+0.28	\\
27524	&		&	+0.11	&		&		&	+0.18	&	+0.59	&	+0.55	&		&	+0.61	&	+0.02	\\
27808	&		&	+0.26	&		&		&	+0.23	&	+0.11	&	+0.59	&		&	+0.34	&	$-$0.16	\\
29391	&		&	+0.00	&		&		&	+0.44	&	+0.46	&	+0.75	&		&	+1.05	&	+0.27	\\
33256	&		&	$-$0.14	&		&		&	$-$0.27	&	$-$0.35	&	+0.40	&		&	+0.19	&	$-$0.35	\\
65123	&	$-$0.09	&	+0.02	&		&		&	$-$0.01	&	$-$0.16	&	+0.27	&	+0.30	&	+0.12	&	$-$0.40	\\
69897	&	$-$0.15	&	$-$0.15	&		&		&	$-$0.21	&	$-$0.33	&	+0.06	&	+0.08	&	+0.01	&	$-$0.52	\\
110093	&		&	$-$0.33	&		&		&	$-$0.08	&	+0.45	&		&	+0.41	&	+0.78	&	$-$0.03	\\
114642	&	$-$0.16	&		&		&		&	$-$0.04	&		&		&	+0.07	&	+0.46	&	$-$0.08	\\
125111	&	$-$0.44	&	$-$0.15	&		&		&	$-$0.35	&	$-$0.53	&	+0.21	&	$-$0.02	&	+0.42	&	$-$0.20	\\
125276	&	$-$0.42	&		&		&		&	$-$0.49	&		&		&	+0.65	&		&		\\
128167	&	$-$0.54	&	$-$0.44	&		&		&	$-$0.44	&	$-$0.56	&	$-$0.01	&	$-$0.20	&	+0.02	&	$-$0.61	\\
135208	&	$-$0.43	&		&		&		&	$-$0.37	&		&		&	$-$0.18	&		&		\\
137006	&	$-$0.13	&	+0.13	&		&		&	$-$0.27	&	+1.08	&	$-$0.24	&	+1.06	&	+0.96	&	+0.15	\\
147449	&	$-$0.05	&	+0.21	&		&		&	$-$0.10	&	+0.68	&	$-$0.03	&		&	+0.91	&	+0.21	\\
157950	&	$-$0.54	&	$-$0.69	&		&		&	$-$0.48	&	$-$1.71	&		&	$-$0.01	&	+0.83	&	+0.21	\\
185124	&	+0.39	&		&		&		&	+0.15	&		&		&	+0.44	&		&		\\
189712	&	$-$0.23	&	$-$0.43	&	$-$0.30	&	$-$0.27	&	$-$0.45	&	$-$0.49	&		&		&		&		\\
191841	&	+0.03	&	$-$0.05	&	$-$0.19	&		&	$-$0.36	&	$-$0.45	&		&		&		&		\\
198390	&	$-$0.31	&	$-$0.21	&	$-$0.06	&	$-$0.09	&	$-$0.32	&	$-$0.35	&	+0.38	&	$-$0.13	&	$-$0.01	&	$-$0.58	\\
199124	&	$-$0.39	&	$-$0.22	&		&		&	$-$0.43	&	$-$1.44	&	$-$0.70	&		&	+0.95	&	+0.20	\\
208362	&	$-$0.18	&	$-$0.50	&	$-$0.06	&	$-$0.66	&	$-$0.48	&	$-$0.61	&	+0.09	&	$-$0.07	&	+0.38	&	$-$0.25	\\
211575	&	+0.23	&	+0.05	&		&	+0.59	&	+0.14	&	$-$0.27	&	+0.67	&	+0.54	&	+0.53	&	$-$0.05	\\
218804	&	$-$0.25	&	+0.00	&		&	$-$0.48	&	$-$0.07	&	$-$0.41	&	+0.67	&	$-$0.12	&	+0.27	&	$-$0.28	\\
222368	&	$-$0.11	&	+0.02	&		&		&	$-$0.36	&	$-$0.23	&	+0.52	&	$-$0.30	&	+0.19	&	$-$0.31	\\
223346	&	$-$0.15	&	$-$0.42	&		&		&	$-$0.20	&	$-$0.21	&	+0.47	&	+0.16	&	+0.32	&	$-$0.21	\\
\hline
\end{tabular}
\end{center}
\end{table*}

To compute model atmospheres for the target stars, we employed the {\sc
LLmodels} stellar model-atmosphere code \citep{Shuly04}.  Local thermodynamical
equilibrium (LTE) and plane-parallel geometry were assumed for all the
calculations. We used the {\sc VALD} database \citep{vald1,vald2,vald3} as a
source of atomic line parameters for the opacity calculations and abundance
determinations.  We also assumed a universal microturbulent velocity
($\upsilon_{\mathrm{mic}}$) of 2.0\,$\mathrm{km\,s}^{-1}$ \citep{Land09}; this
choice was confirmed by our analysis of \ion{Fe}{I} lines for the few stars
for which our spectra included enough lines that were both weak and strong and
could thus support a reliable determination of $\upsilon_{\mathrm{mic}}$.

We derived the LTE abundances of the low-\ensuremath{{\upsilon}\sin i} stars
(\ensuremath{{\upsilon}\sin i}\,$\leq$\,30\,$\mathrm{km\,s}^{-1}$) by analysing
measured equivalent widths with a modified version \citep{Tsym96} of the {\sc
WIDTH9} code \citep{Kuru93}.  For the faster rotators we derived LTE abundances
by fitting synthetic spectra that had been calculated with {\sc SYNTH3}
\citep{Koch07}.  The tools and methodology adopted were the same as those
described in \citet{Foss07}. For solar abundances we used the values from
\citet{Grev96} in order to be compatible with the abundances published for the
$\lambda$ Bootis stars by \citet{Paun02}. The uncertainties in the
abundances were typically $\pm$0.2\,dex.

The oxygen triplet at 7771-5\,\AA\, is known to be severely influenced by
non-LTE (NLTE) effects \citep{Paun99}. We therefore used the corrections given
by \citet{Sitn13} for $\log g$\,=\,4.0 (see their Fig. 5) as they are valid
for main-sequence stars.  We applied the NLTE corrections according to the
effective temperatures listed in Table \ref{parameters}.  
None of the other lines was significantly affected by NLTE.  In the
case of carbon, for instance, NLTE effects for the lines that we observed are
below 0.1\,dex \citep{Paun99}. The abundances thus derived are listed in Table
\ref{abundances}.

We checked our results against those from the Geneva-Copenhagen survey
\citep{Holm09} and the PASTEL catalogue \citep{Soub10}. The first includes
kinematic data, effective temperature and absolute magnitudes, whereas the
second includes effective temperature, surface gravity and [Fe/H].
The differences between the values are listed in Table \ref{diffs}. We found
excellent agreement between all the parameters with those listed in the
literature, thus strengthening confidence in our reduction, calibration and
analysis methods.

The element abundances of superficially normal stars were taken from
\citet{Adel91a, Adel91b, Adel94, Adel96, Adel97, Cali97, Hill93, Hill95,
Take09, Vare99}.  The selected objects cover the complete range of
\ensuremath{T_{\mathrm{eff}}} and \ensuremath{\log g} in the other two samples.
Values for the $\lambda$ Bootis stars were taken from \citet[][Table 1
therein]{Paun02}.  We checked the literature, but did not find reports of any
additional light-element abundances for the stars listed there.

\begin{figure}
\begin{center}
\includegraphics[width=75mm]{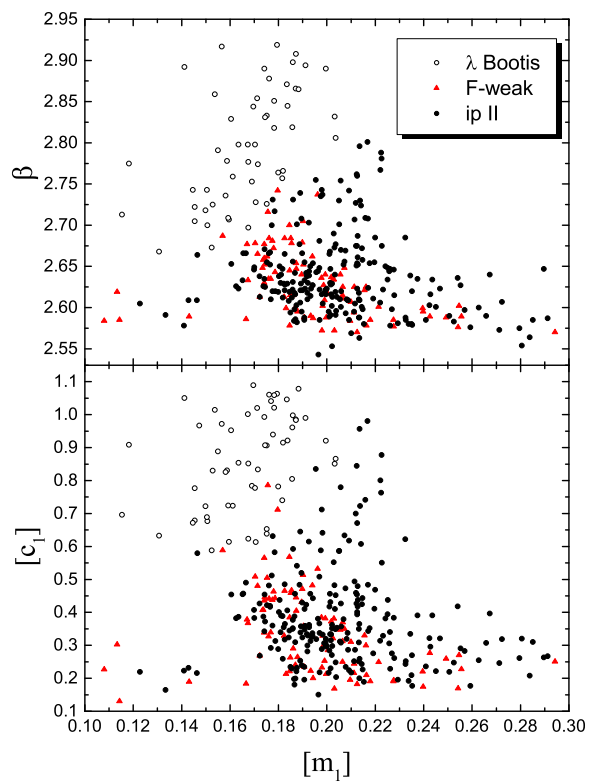}
\caption{Photometric plots of the unreddened indices $[m_1]$, $[c_1]$,
and $\beta$ for $\lambda$ Bootis (open circles), F-weak (red triangles),
and ip\,II (black triangles) stars.} 
\label{photometry}
\end{center}
\end{figure}

\begin{figure}
\begin{center}
\includegraphics[width=75mm]{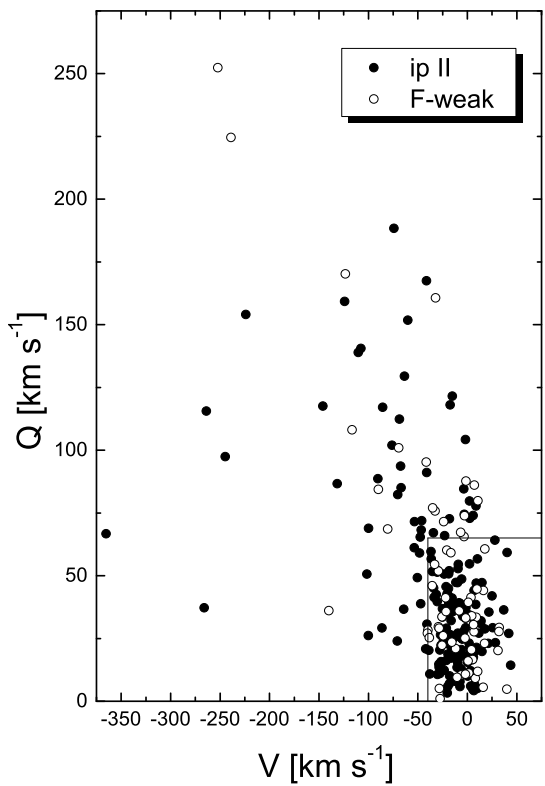}
\caption{The $Q$\,=\,$\sqrt{U^2+W^2}$ versus $V$ diagram for ip\,II (filled circles) and F-weak (open circles) stars.  The plot is a measure of the 
kinematic energy not associated with rotation.  The two groups
are not distinguishable. The rectangle marked is the
empirical limit for disk stars according to \citet{Calo99}.}
\label{motion_1}
\end{center}
\end{figure}

\begin{table}
\caption{The differences between our derived values for the target sample and
those in the  Geneva-Copenhagen survey \citep[GCS, ][]{Holm09} and the PASTEL catalogue \citep{Soub10}.}
\label{diffs}
\begin{center}
\begin{tabular}{lcccc}
\hline
\hline 
Parameter & \multicolumn{2}{c}{GCS} & \multicolumn{2}{c}{PASTEL} \\
          & mean & N & mean & N \\
\hline
$\Delta$\ensuremath{T_{\mathrm{eff}}} [K] & +44$\pm$54 & 33 & +8$\pm$78 & 35 \\  
$\Delta$\ensuremath{\log g} [dex] & & & +0.03$\pm$0.14 & 22 \\
$\Delta$\ensuremath{M_{\mathrm{V}}} [mag] & $-$0.01$\pm$0.08 & 33 & & \\
$\Delta$[Fe/H] [dex] & & & +0.05$\pm$0.14 & 22 \\
$\Delta U$ [km\,s$^{-1}$] & +0.33$\pm$1.32 & 33 & & \\
$\Delta V$ [km\,s$^{-1}$] & $-$0.61$\pm$2.11 & 33 & & \\
$\Delta W$ [km\,s$^{-1}$] & +0.51$\pm$1.48 & 33 & & \\
\hline
\end{tabular}
\end{center}
\end{table}

\section{Discussion} 

There is considerable confusion in the literature over the definitions
and origins of weak-lined F-type stars on the one hand and ip\,II stars on the
other, and this question needs further discussion.

\subsection{The weak-lined F-type stars and the ip\,II stars} 
\label{ipii_fweak}

We first summarize the results that have already been published, and then 
analyse the available photometric and kinematic data.

\subsubsection{Historical facts}

Baade's original concept of a stellar `population' in the Milky Way was based
on a disk (I) and a halo (II) population.  The concept later evolved into a
scheme with subdivisions.  In one such subdivision, the old metal-weak
Population II was divided into halo (extreme) and ip\,II objects.  The latter
was comprised of objects that had a velocity dispersion, a chemical composition,
and a concentration towards the Galactic plane that was intermediate between
those of the halo and the disk populations.  The ip\,II population was later
proposed in order to explain and fit both the global and the local Galactic
kinematics. \citet{Robi86} estimated that about 1.5\,\% of all stars
in the solar neighbourhood are ip\,II objects.

However, \citet{Stro64} and \citet{Blau65} had already pointed out that, owing
to a steady transition between the populations and also to insufficient precision 
in their ages, no clear distinction between the groups could be
made. \citet{Stro66} therefore proposed using the $uvby\beta$ photometric
system, as it would be more illuminating.  The $\delta m_1$ index, for instance,
was used to distinguish solar from underabundant objects in the spectral-type
domain between early F and mid G. \citet{Olse79} listed the following boxes
for cool ip\,II type stars (i.e. later than F4):

\begin{eqnarray}
+0.270<&\left(b-y\right)&<+0.412, \\
+0.045<&\delta m_1&<+0.100, \\
-0.045<&\delta c_1&<+0.280.
\end{eqnarray} 

For the hotter stars (+0.043\,$<$\,$(b-y)$\,$<$\,+0.270 and
+0.045\,$<$\,$\delta m_1$) Str\"omgren found overlaps with `classical'
$\lambda$ Bootis stars and with field horizontal-branch stars which could then
be separated easily by their $c_1$ values.

\begin{figure*}
\begin{center}
\includegraphics[width=160mm]{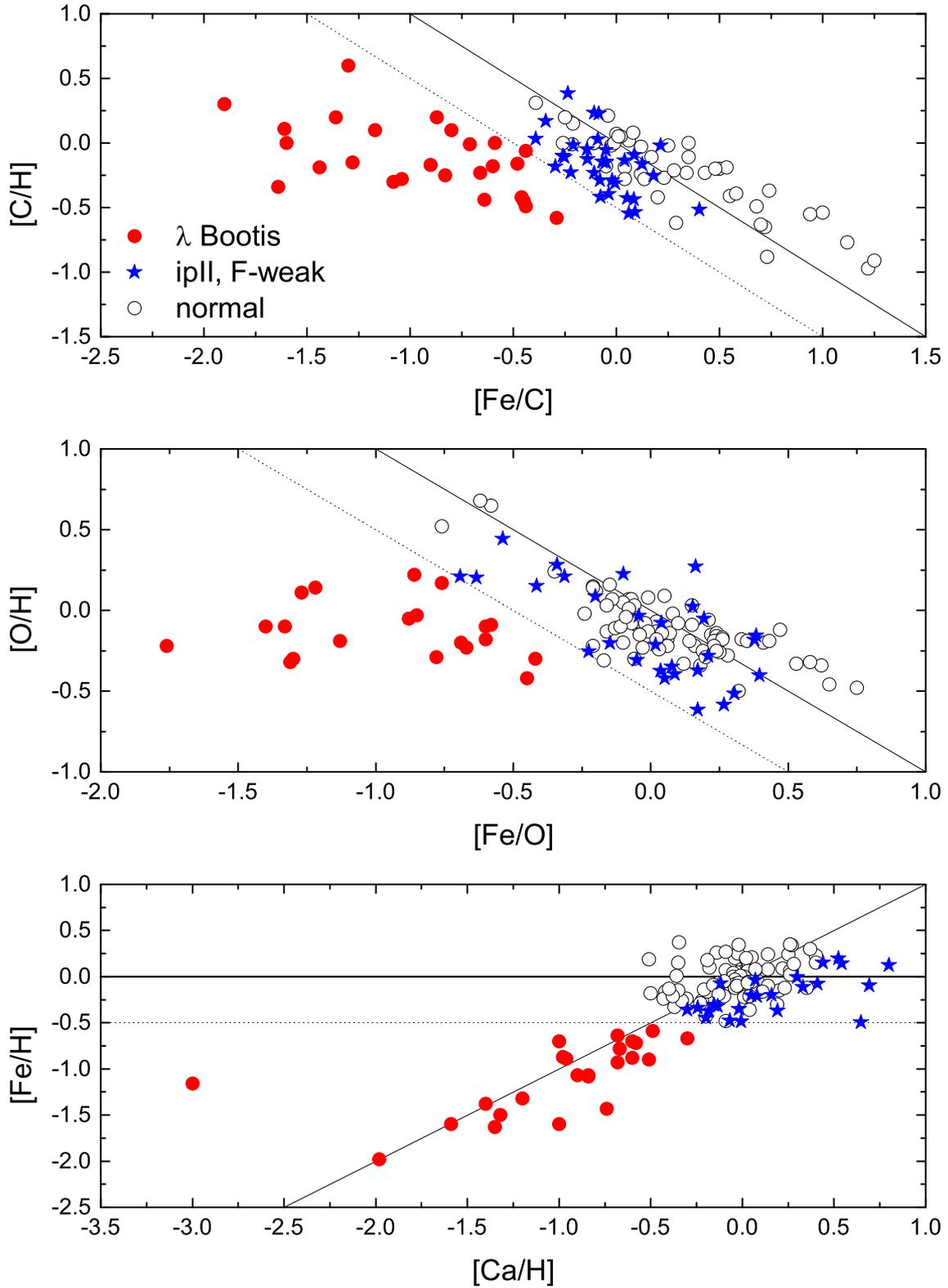}
\caption{Plots of different element abundances for the star groups 
investigated. The $\lambda$ Bootis stars are clearly separated.}
\label{abundances_plot}
\end{center}
\end{figure*}

\citet{Gray89} obtained spectra at a classification dispersion
(67\,\AA\,mm\,$^{-1}$) of a sample of metal-weak stars that were selected
according to the above criteria.  He established metal-weak standards for that
spectral range so as to extend the Yerkes MK system, and labelled his sample
``intermediate Population II-type stars''.

Independently, \citet{Jasc89} conducted a survey of metal-weak F-type stars,
and also used spectra with a classification dispersion
(80\,\AA\,mm\,$^{-1}$). Both Jaschek et al.~and Gray employed the strengths of
the Balmer lines, the G-band (CH), and metallic lines, but Jaschek et
al.~labelled their sample of metal-weak F-type stars simply ``F-weak type
stars''. \citet{Hauc91} subsequently investigated the same sample as Jaschek et al.
using the Geneva photometric system; the
metallicity-sensitive indices that calibrate [Fe/H] are the same as those in
the Str{\"o}mgren system.

In this paper we have therefore adopted the expression ip\,II group/stars to
describe stars listed in the literature as members of either the ip\,II or the
F-weak group. 

\subsubsection{Photometric and kinematic data} 

We employed the Str{\"o}mgren $uvby\beta$ system search for similarities
between the two groups.  Figure \ref{photometry} plots unreddened indices
$[m_1]$, $[c_1]$, and $\beta$ for $\lambda$ Bootis stars \citep{Paun02}, F-weak
\citep{Jasc89} stars, and ip\,II stars \citep{Gray89}. To derive reddening-free
parameters we used the definition originally introduced by \citet{Stro66}:
\begin{eqnarray}
\left[\mathrm{m_1}\right]&=&\mathrm{m_1}+0.18\left(b-y\right), \\
\left[\mathrm{c_1}\right]&=&\mathrm{c_1}-0.20\left(b-y\right).
\end{eqnarray} 
In general, $\beta$ is temperature sensitive, $[m_1]$ correlates with
metallicity, and $[c_1]$ is correlated with surface gravity and thus with the
age of a star for a given spectral type. As the figure indicates, there is no
clear separation between the two groups.

That result from the photometric data was supported by kinematical data which
were computed from proper motions (from the {\it Hipparcos} catalogue) and mean
radial velocities (from the literature) as described in \citet{John87}. A
right-handed coordinate system for $U$, $V$, and $W$ was used, so the velocities
are positive in the directions of the Galactic centre, Galactic rotation and
the North Galactic Pole. The standard solar motion in that system is in units 
of [km\,s$^{-1}$] (+10.4, +14.8, +7.3) according to \citet{Miha68}. For our analysis we
adopted the $V$, and the quantity $Q$\,=\,$\sqrt{U^2+W^2}$ which is a measure
of the kinematic energy other than that associated with rotation. Since
all velocities are heliocentric, stars with $V$\,$>$\,$-$40\,km\,s$^{-1}$ and
$Q$\,$<$\,65\,km\,s$^{-1}$ are probable disk members \citep{Calo99}. Figure
\ref{motion_1} plots $Q$\,=\,$\sqrt{U^2+W^2}$ against $V$.
There is no distinction between the two groups in that 
diagram.

The following conclusion can be drawn from this analysis:
\begin{itemize}
\item There are no obvious differences between the sample chosen by
\citet{Gray89} and the one selected by \citet{Jasc89}.  This means that 
ip\,II and F-weak type stars are simply names for the same group of stars.
\end{itemize}

\begin{figure}
\begin{center}
\includegraphics[width=75mm]{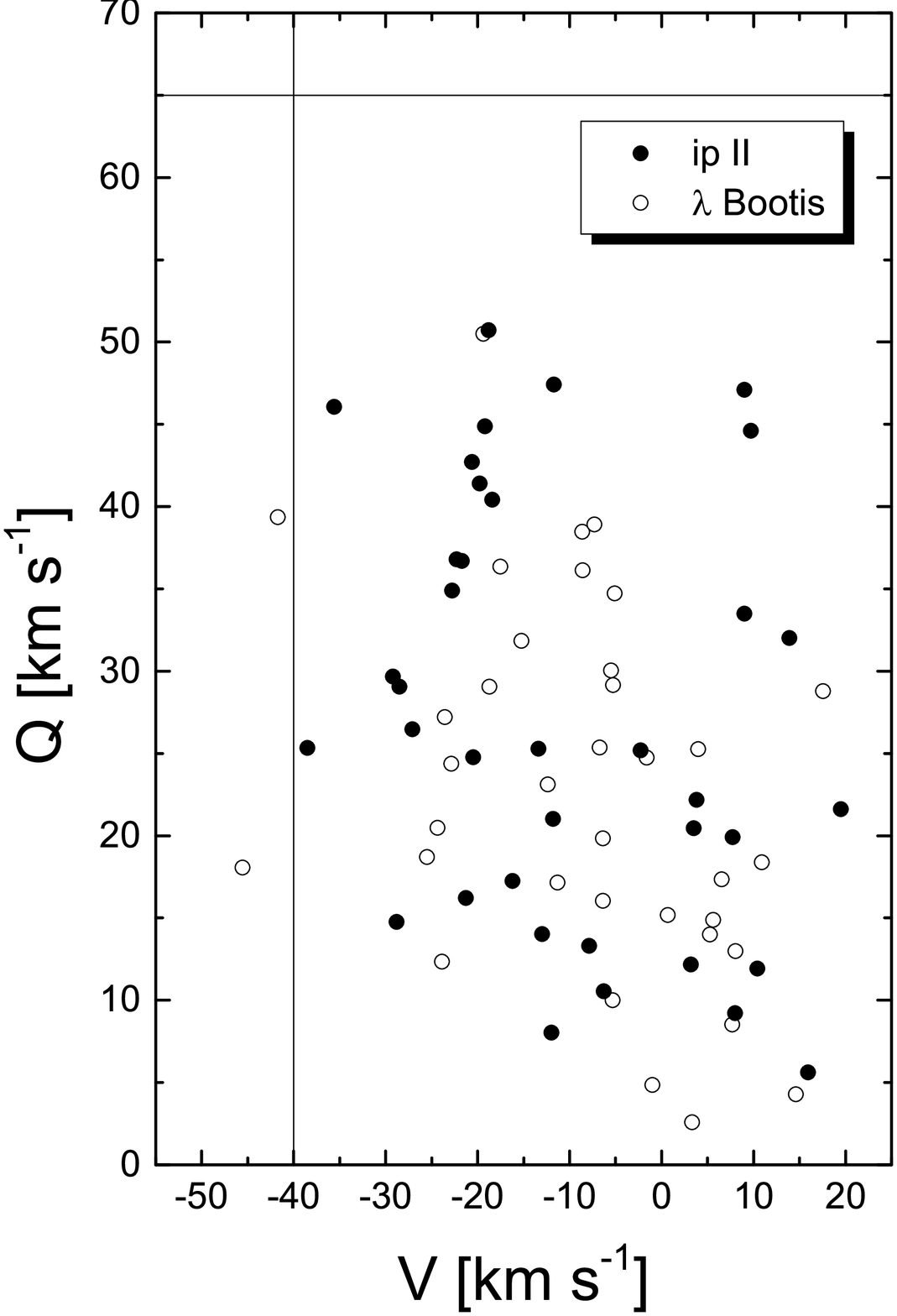}
\caption{$Q$\,=\,$\sqrt{U^2+W^2}$ plotted against $V$.  $\lambda$ Bootis 
stars are shown as open circles, ip\,II stars as filled circles.
The horizontal and vertical lines are the empirical limits for 
disk stars according to \citet{Calo99}.}
\label{motion_2}
\end{center}
\end{figure}

\subsection{The groups of $\lambda$ Bootis and ip\,II stars}

One of the most important features that distinguishes $\lambda$ Bootis from other stellar
groups is their unusual abundance patterns. Although the light elements have solar
abundances, the heavier elements are overabundant but nearly to the same extent.
This means that, for example,
silicon shows more or less the same amount of overabundance as iron.
We therefore investigated these abundance
characteristics shown by our selected sample.  Figure \ref{abundances_plot}
shows the results for three different combinations: [Fe/C] versus [Fe/H],
[Fe/O] versus [Fe/H], and [Fe/H] versus [Ca/H]. In all three diagrams, the
ip\,II stars are located among the normal non-peculiar objects,  the
$\lambda$ Bootis stars are clearly separated from them, and the light elements 
are offset by $-$0.5 dex. The behaviour of the heavier elements was then
examined for possible correlations between the samples in respect of
[Fe/H] versus [Ca/H]. For the normal, ip\,II, and $\lambda$ Bootis samples
(see Fig. \ref{abundances_plot}) we derived

\begin{eqnarray}
\left[\mathrm{Fe/H}\right]&=&-0.01\left(2\right)+0.32\left(9\right)\left[\mathrm{Ca/H}\right],N=95, \\
\left[\mathrm{Fe/H}\right]&=&-0.25\left(4\right)+0.26\left(9\right)\left[\mathrm{Ca/H}\right],N=26,\\
\left[\mathrm{Fe/H}\right]&=&-0.68\left(13\right)+0.40\left(11\right)\left[\mathrm{Ca/H}\right],N=24, \\
\left[\mathrm{Fe/H}\right]&=&-0.32\left(11\right)+0.83\left(12\right)\left[\mathrm{Ca/H}\right],N=23.
\end{eqnarray}
here $N$ denotes the number of available individual abundances for each star group.

The last correlation did not include the outlier (HD~91130) which has a calcium
abundance of $-$3.0\,dex.  The first two samples show a slope of about 0.30
whereas the $\lambda$ Bootis stars show almost no slope. The ip\,II sample
overlaps the normal stars across a wide range of abundances.

From this analysis, we define a $\lambda$ Bootis star on the basis of elemental
abundances if
\begin{enumerate}
\item {[C/H]}\,$<$\,$-$0.50\,$-$\,[Fe/C],
\item {[O/H]}\,$<$\,$-$0.50\,$-$\,[Fe/O],
\item {[Fe/H]}\,$<$\,$-$0.50,
\item Heavy elements scale with [Fe/H].
\end{enumerate}
This definition separates the ip\,II stars from $\lambda$ Bootis stars. 

We also analysed the photometric and kinematic data of both groups.  (see
Sect. \ref{ipii_fweak}).  Figure \ref{photometry} plots the unreddened indices
$[m_1]$, $[c_1]$, and $\beta$. The $\lambda$ Bootis stars scarcely overlap at
all with the ip\,II objects. That characteristic could become an additional
criterion for distinguishing both groups.

Figure \ref{motion_2} shows the kinematic data for the $\lambda$ Bootis stars
and the ip\,II (Table \ref{parameters}) sample.  The two outliers (HD~6870 and
HD~84123) were already been discussed in \citet{Paun02}. Their $U$ and
$W$ velocities qualify them as being Population I objects. In respect of our
total sample, we can state that they all belong kinematically to the Galactic
disk. We therefore conclude that there is no distinction between the vast
majority of the members of either group on the basis of their kinematic data.

\begin{table}
\caption{[Z] values for the $\lambda$ Bootis stars \citep[from][]{Paun02} using
individual elemental abundances. The absolute solar values were taken from
\citet{Grev96}; [Z]$_{\odot}$\,=\,0.017.}
\label{z_values}
\begin{center}
\begin{tabular}{lc|lc}
\hline
\hline 
Star & [Z] & Star & [Z] \\
\hline
HD~319	&	0.018	&	HD~120500	&	0.013	\\
HD~11413	&	0.010	&	HD~125162	&	0.007	\\
HD~15165	&	0.008	&	HD~142703	&	0.007	\\
HD~31295	&	0.007	&	HD~168740	&	0.010	\\
HD~74873	&	0.024	&	HD~170680	&	0.012	\\
HD~75654	&	0.014	&	HD~183324	&	0.011	\\
HD~84123	&	0.009	&	HD~192640	&	0.007	\\
HD~91130	&	0.018	&	HD~193256	&	0.008	\\
HD~101108	&	0.016	&	HD~193281	&	0.012	\\
HD~106223	&	0.024	&	HD~198160	&	0.010	\\
HD~107233	&	0.006	&	HD~204041	&	0.006	\\
HD~110411	&	0.017	&	HD~210111	&	0.008	\\
HD~111604	&	0.007	&	HD~221756	&	0.018	\\
\hline
\end{tabular}
\end{center}
\end{table}

\subsection{Evolutionary status and metallicity}

\begin{figure}
\begin{center}
\includegraphics[width=75mm]{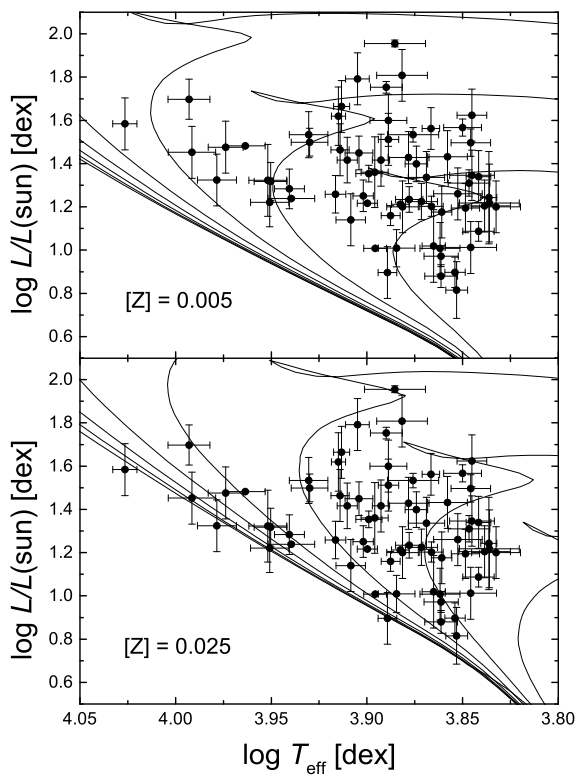}
\caption{$\lambda$ Boo stars in the H--R diagram, with isochrones
from \citet{Bres12} for the whole range of [Z] values listed in Table \ref{z_values}.} 
\label{isochrones}
\end{center}
\end{figure}

In order to investigate the possibility of an intrinsic chemical peculiarity in
the $\lambda$ Bootis stars, we calculated individual [Z] values for 26 objects
whose abundances are listed in Table 5 of \citet{Paun02}. The absolute mass
fraction abundances ($\chi$) were taken from \citet{Grev96};
[Z]$_{\odot}$\,=\,0.017.  Where light-element abundances were not known, the
value for carbon was used; the value for iron was adopted as representative of
Fe-peak elements.  In total, we included all elements for which
$\chi$\,$>$\,10$^{-8}$.  Table \ref{z_values} lists our results for the sample.
The light elements (C, N, O, and S) distribute the most, in absolute
numbers, for [Z]. Therefore, two stars (HD~74873 and
HD~106223) even have absolute overabundances with [Z]\,=\,0.024, while all
other objects have values as low as 0.006.  As a next step, we examined the
relationship between metallicity and ages by selecting isochrones from \citep[from][]{Bres12} for the
corresponding metallicity values. Figure \ref{isochrones} plots the
Hertzsprung-Russell diagram for the corresponding range of [Z]. Several stars lie beyond the
terminal-age main sequence (luminosity class IV), but none close to the zero-age main-sequence. The
complete metallicity range is compatible with the observed data, so (within the
errors) no objects is located below the zero-age main sequence (which from an
astrophysical standpoint would not be possible). 

\section{Conclusions and outlook}

We investigated a possible connection between the $\lambda$ Bootis stars and
intermediate Population II, F-weak stars on the basis of elemental abundances
and kinematic data.  It tackled the important questions whether the $\lambda$
Boo stars are intrinsically metal-weak, and thus belong to the intermediate
Population II, or whether their ususual abundances are restricted to the outer
layers of the atmosphere. Recent results from an asteroseismologic study
suggest the former.  If so, the $\lambda$ Bootis stars become even more
interesting, especially those in which accretion disks have been detected.

Our analysis of 38 ip\,II/F-weak stars was based on detailed element
abundances, including the light elements carbon and oxygen, and kinematic data
from the {\it Hipparcos} catalogue.

First we investigated whether the ip\,II and F-weak stars are separate groups
or whether the names are merely synonyms for identical objects. From
photometric, spectroscopic and kinematic data, we conclude that the two groups
are indeed identical.

We then compared the elemental abundances of normal, $\lambda$ Bootis, and
ip\,II stars.  The ip\,II group overlaps with the normal stars, but is completely separated from the $\lambda$ Bootis objects. We, therefore, defined four
constraints regarding the discrimination of both groups. From the kinematic data,
no similar differentiation can be drawn.

Finally, we calculated the overall [Z] values for the $\lambda$ Bootis stars
and plotted them, along with corresponding isochrones, in the H--R
diagram. Using metal-poor isochrones would shift several stars beyond the
terminal-age main sequence.

As independent tests, we propose that asteroseismological studies \citep{Casa09} be made of
pulsating members, and that spectroscopic binary systems \citep{Ilie02} containing $\lambda$
Bootis-type components be investigated. One very interesting star in that
respect is HD~210111, which is a $\delta$ Scuti pulsator in a spectroscopic
binary system. 

Another interesting stellar system is the eclipsing binary HR~854 ($\tau$ Persei). 
The primary is a late G-type giant, and the secondary could belong to the $\lambda$ Bootis 
group \citep{Grif92}. Its K-line type is earlier than derived from the hydrogen lines \citep{Gine02}.
Therefore, this system is an excellent test for fitting isochrones of different metallicities.

\begin{acknowledgements}
This project is financed by the SoMoPro II programme (3SGA5916). The research leading
to these results has acquired a financial grant from the People Programme
(Marie Curie action) of the Seventh Framework Programme of EU according to the REA Grant
Agreement No. 291782. The research is further co-financed by the South-Moravian Region. 
It was also supported by the grant 7AMB14AT015,
the financial contributions of the Austrian Agency for International 
Cooperation in Education and Research (BG-03/2013 and CZ-09/2014), NSF grant 01/7 - DNTS AT, and the Austrian Research 
Fund via the project FWF P22691-N16. We thank our referee, Elizabeth Griffin, for her efforts
to significantly improve this paper.
This work reflects only the author's views and the European 
Union is not liable for any use that may be made of the information contained therein.
\end{acknowledgements}

\end{document}